\documentclass[prb,twocolumn,floatfix,showpacs,preprintnumbers, 
               superscriptaddress,amsmath,amssymb]{revtex4}
\usepackage[dvips]{graphics}
\usepackage{psfrag}
\usepackage{graphicx}
\usepackage{epsf}
\usepackage{subfigure}
\usepackage{xspace}
\usepackage{ucs}
\usepackage[utf8x]{inputenc}
\usepackage{amssymb}
\newcommand{\be}{\begin{eqnarray}}
\newcommand{\ee}{\end{eqnarray}}
\newcommand{\iv}{$I$-$V$\xspace}
\newcommand{\ie}{i.\,e.\ }
\newcommand{\eg}{e.\,g.\ }

\begin{document}
\title{Charge correlations in polaron hopping through molecules}

\author {Benjamin B. Schmidt} 
\affiliation{Institut f\"ur Theoretische Festk\"orperphysik
and DFG-Center for Functional Nanostructures (CFN),
Universit\"at Karlsruhe, 76128 Karlsruhe, Germany}
\affiliation{Forschungszentrum Karlsruhe, Institut f\"ur Nanotechnologie,
  Postfach 3640, 76021 Karlsruhe, Germany}
\author {Matthias H. Hettler} 
\affiliation{Forschungszentrum Karlsruhe, Institut f\"ur
  Nanotechnologie, Postfach 3640, 76021 Karlsruhe, Germany}
\author {Gerd Sch\"on} 
\affiliation{Institut f\"ur Theoretische Festk\"orperphysik
and DFG-Center for Functional Nanostructures (CFN),
Universit\"at Karlsruhe, 76128 Karlsruhe, Germany}
\affiliation{Forschungszentrum Karlsruhe, Institut f\"ur
  Nanotechnologie, Postfach 3640, 76021 Karlsruhe, Germany}

\date{\today}
\begin{abstract}
In many organic molecules the strong coupling of excess charges
to vibrational modes leads to the formation of polarons, i.e., a localized 
state of a charge carrier and a molecular deformation. Incoherent hopping of polarons along the molecule is the dominant mechanism of transport at room temperature. We study the far-from-equilibrium situation where, due to the applied bias, the induced number of charge carriers on the molecule is high enough such that charge correlations become relevant. We develop a diagrammatic theory that exactly accounts for all many-particle correlations functions for incoherent transport through a finite system. We compute the transport properties of short sequences of DNA by expanding the diagrammatic theory up to second order in the hopping parameters. The correlations qualitatively modify the \iv characteristics as compared to those approaches where correlations are dealt with in a mean-field type approximation only.

\end{abstract}

\maketitle

\section{Introduction} 
\label{sec:Introduction}
Molecular electronics experiments performed during recent years have probed the 
conductance and current-voltage characteristics of a large variety of molecules. 
Several experiments on long molecules indicate that transport is is not described by coherent Landauer transport or tunneling
but rather by an incoherent hopping of charge carriers along the molecule. Examples are experiments on DNA\cite{Yoo01,Shigematsu03} or oligophenyleneimine wires\cite{Choi08}. 
In the latter experiment the length dependence of the conductance clearly demonstrated 
the crossover from the coherent (tunneling) to the incoherent transport regime at a molecule 
length of about $4 nm$.

In many experiments,
the molecule consists of repeated segments (either identical or with chemical modifications) where the 
quantum mechanical hopping amplitude between the segments can be tuned to some 
degree by the choice of the ``linker group''. If the hopping amplitude between the segments is large (e.g., for stiff molecules with fully conjugated electron systems) the quantum-mechanical coherence on the molecule can extend quite far at low temperatures, such that even molecules of up to several $nm$ length display signs of coherent transport, at least within the molecule\cite{kubatkin03,osorio07}. 
On the other hand, if the coupling of 
segments is weak (e.g., for flexible molecules with weakly conjugated or saturated linker groups) the coherence decays quickly, such that charge carriers are typical localized over a single or a few segments only. In this case, the molecule tends to change its conformation
in order to lower its energy when charged, a process called polaron formation. The polaron is a combination of a charge carrier and the localized deformation. At room temperature, charge transport is then dominated by incoherent hopping of polarons along the molecule. At low temperature, coherent ``band-like'' transport of polarons might be observable.

The theoretical description of polaronic effects in molecule-electrode setups so far 
either focused on molecular single-level systems~\cite{Galperin04,Alexandrov07,Alexandrov06,Mitra05} or described polaron hopping in long molecules by assuming `simple' rate equations~\cite{Berlin01,Schmidt08}.
B\"ottger and Bryksin have shown in a rigorous description of polaron transport in
bulk systems  that even in the absence of Coulomb interactions phonon-mediated charge correlations between different sites develop~\cite{Boettger85}.
However, in their calculations they included these correlations only in a mean-field like
manner. The earlier rate equation approaches to polaron hopping in nanoscale systems also treat correlations within this mean-field approximation.

The mean-field approximation to many-particle correlations can be justified in systems with 
low density of charge carriers, and is a sufficient approximation for many doped (organic) semiconductors.
In molecular electronics experiments, however, where a transport bias 
on the order of $1\,\rm V$ is applied over a molecule of a few nanometer length, the
average charge density may be much higher, and correlations become relevant. 
For example, for small molecules often the Coulomb interaction (or charging energy) 
dominates, leading at low temperatures to transport characteristics similar to single-electron transistors.\cite{park00,Grabert92} 
With increasing molecule size the relevance of Coulomb blockade decreases, but still, the transport {\it along} the molecule is affected by charge correlations, either due to (non-local) Coulomb interaction or the (retarded) interactions mediated by the coupling of charge carriers to vibrational modes mentioned above. We will demonstrate that in general such correlations are not sufficiently described by a mean-field approach.

We have extended the diagrammatic approach by Boettger and Bryksin 
to describe molecular systems coupled to metallic electrodes. In the usual diagrammatic 
approach to small molecules (or quantum dots)\cite{Koenig96,hettler_prl} the molecular 
{\it eigenstates} including the Coulomb interaction are the basis of a perturbative expansion in the weak coupling to metallic electrodes. In contrast, in the present problem the ``basis states'' are ``local'' to the molecule segments (e.g. a DNA base pair or a phenylene ring). The expansion parameters also include the small hopping amplitudes between the molecular segments. As the perturbation expansion usually involves a ``self-energy'' resummation, where diagrams of a certain type (but of arbitrarily high order) are accounted for, the occupation numbers (or, in general, one-particle correlation functions) are coupled to higher order correlations functions, unless the system is non-interacting. Similar to, e.g., equation-of-motion methods, a hierarchy of equations for the correlation functions can be generated, which has to be truncated in some manner (often following more numerical necessities than physical arguments). 

The present paper demonstrates that an exact description of correlation effects 
mediated by vibrational modes is possible for a {\it finite size system}. This is because the hierarchy is naturally truncated at the level of the highest possible correlation 
function involving all `sites' of the molecule. The resulting finite set of coupled {\it linear} 
equations for the occupation number and the many-particle correlation functions can then be solved without resorting to any further truncation procedure.

As an example we study short DNA molecules coupled to metallic electrodes.
The DNA is modeled by a tight-binding chain identifying
each base pair with a single tight-binding site. The sites describing either 
guanine-cytosine (GC) or adenine-thymine (AT) base pairs have different onsite energies and
are coupled by direction- (sequence-) dependent hopping integrals $t_{ij}$ (compare Table~\ref{tab:hopping}).
The polarons are formed by strong coupling of the charge degrees of freedom to local
base pair vibrations. To ensure energy dissipation, these base pair vibrations are in turn
coupled to a set of harmonic oscillators, describing the influence of a dissipative environment.

In the {\sl first part} of this article we introduce the diagrammatic technique 
describing incoherent polaron hopping transport through molecules or other nanoscale systems 
which are coupled to metallic electrodes. This technique allows the description of polaron 
transport with the exact consideration of correlation effects arising from the electron-vibration interaction. The approximation of the technique lies in the need to restrict
the order of the expansion in small hopping parameters. 
In the {\sl second part} we apply this diagrammatic technique to polaron hopping transport through short DNA molecules coupled to metallic electrodes with the following results: (i) Correlations effects beyond the mean field approximation are relevant for the linear conductance in inhomogeneous DNA molecules already for low charge densities of less than 1 percent. (ii) When a transport bias is applied over the molecule, correlation effects become important even when the occupation in equilibrium is negligible. (iii) Inhomogeneous DNA molecules in general exhibit two maxima in the zero bias conductance as a function of equilibrium chemical potential (gate voltage) and also in the  differential conductance as a function of applied transport bias. In contrast, the mean field approach only displays one maximum. (iv) Depending on the sequence, the secondary maxima can be suppressed, as a consequence of a small hopping rate limiting the transport through the system. Details of the diagrammatic technique are shown in the Appendices~\ref{app:diag_rules}-\ref{app:op_prod}.  

\section{Model and Technique}
\label{sec:model_and_technique}
The minimal Hamiltonian to describe polaron transport through DNA is 
$H=H_{\rm el}+H_{\rm vib}+H_{\rm el-vib}+H_{\rm L}+H_{\rm R}+H_{\rm T,L}+H_{\rm T,R}+H_{\rm bath}$
with
\begin{align}
H_{\rm el} &=& \sum_i \hat{\epsilon}_i a_i^{\dagger}a_i -\sum_{<ij>} t_{ij} a_i^{\dagger}a_j \nonumber\\
H_{\rm T,L}+H_{\rm T,R} &=& \sum_{n,r,i} \left[ t^{r}c_{nr}^{\dagger}a_i+t^{r}a_i^{\dagger}c_{n r} \right]\nonumber\\
H_{\rm vib}&=& \sum_{\alpha}\sum_i \omega_{\alpha i}\left(B_{\alpha i}^{\dagger}B_{\alpha i}+\frac{1}{2}\right) \nonumber\\
H_{\rm el-vib} &=& \sum_{\alpha} \sum_i \lambda_{\alpha i} \, a_i^{\dagger}a_i (B_{\alpha i}+B_{\alpha i}^{\dagger})\; .
\end{align}

The term $H_{\rm el}$ models the electrons on the molecule with operators 
$a_{i}^{\dag}, a_{i}$ in a single-orbital tight-binding representation. This implies 
that the molecule consists of $N$ parts (labeled $i$). The electronic properties of the 
molecule can then be described by the molecular orbitals (usually the HOMO or LUMO) of 
these sub-entities with on-site energies $\epsilon_i$ and hopping $t_{ij}$ between 
neighboring parts of the molecule. The terms $H_{\rm L/R}$ refer to the left and right 
electrodes. They are modeled by non-interacting electrons, described by operators 
$c_{n\, \rm L/R}^{\dag}, c_{n\,\rm  L/R}$, with a flat density of states $\rho_e$ 
(wide band limit). Since we do not focus on the details of the coupling between the 
molecule and the electrodes, it is sufficiently described by $H_{\rm T,L}+H_{\rm T,R}$. 
The tunneling amplitudes are assumed to be independent of the molecular orbital $i$ and 
the quantum numbers of the electrode states $\nu$. The coupling strength is then 
characterized by the parameter $\Gamma^{\rm L,R}\propto \rho_e |t^{\rm L,R}|^2$.

Polarons are formed due to strong coupling of electronic and vibrational 
degrees of freedom. The vibrations labeled $\alpha,i$ are described in $H_{\rm vib}$, 
with bosonic operators $B_{\alpha i}$ and $B_{\alpha i}^{\dagger}$ for the vibrational mode 
with frequency $\omega_{\alpha i}$, \ie every part of the molecule can vibrate independently.
$H_{\rm el-vib}$ couples the electrons on the molecule to the vibrational modes, 
where $\lambda_{\alpha i}$ is the strengths for the local electron-vibration coupling for the 
site $i$ and mode $\alpha$, respectively. To ensure energy dissipation and thermal 
occupation of the vibrational states we couple every vibration to its own bath $H_{i,\rm bath}$,
the microscopic details of which do not matter.

A perturbative treatment of the strong electron-vibration coupling in the above Hamiltonian 
is not reasonable. Nevertheless, to allow for a perturbation expansion we apply the 
so-called polaron or Lang-Firsov unitary transformation 
\be\tilde{H}=e^{S}He^{-S}
\ee
with the generator 
\be
S=-\sum_{\alpha  i}
\frac{\lambda_{\alpha i}}{\hbar \omega_{\alpha i}} \, a_i^{\dagger}a_i
\left[B_{\alpha i}-B_{\alpha i}^{\dagger} \right]\, .
\ee
We introduce transformed electron and vibrational operators,
\be
\tilde{a}_i  &=&  a_i\chi_i \nonumber\\
\tilde{B}_{\alpha i} &=& B_{\alpha i}-\frac{\lambda_{\alpha i}}{\hbar \omega_{\alpha i}} \, a_i^{\dagger}a_i \nonumber
\ee
and polaron operators 
\be
\chi_i &=& \exp \left[\sum_{\alpha} \frac{\lambda_{\alpha i}}{\hbar \omega_{\alpha i}} \,
  (B_{\alpha i}-B_{\alpha i}^{\dagger}) \right]\, .
\ee
Operators $\chi_i$ with different indices $i$ act on different 
vibrational states, therefore they commute for all times.
In terms of these quantities the Hamiltonian  reads
\begin{align}
\tilde{H} =& \tilde{H}_0 + \tilde{H}'   \nonumber\\
\tilde{H}_0=& \sum_i (\epsilon_i-\Delta_i) a_i^{\dagger}a_i+  \sum_{\alpha i} \hbar \omega_{\alpha i} \left(B_{\alpha i}^{\dagger}B_{\alpha i}+\frac{1}{2}\right)\nonumber \\
&+H_{\rm L} +H_{\rm R} \nonumber\\
\tilde{H}'=&-\sum_{<ij>} t_{ij} \, a_i^{\dagger} \chi_i^{\dagger} a_j \chi_j \label{eq:Hperturb2} \nonumber\\
&+ \sum_{\nu,r,i} \left[ t^{r} c_{\nu r}^{\dagger} a_i \chi_i +t^{r*}a_i^{\dagger} \chi^{\dagger}_i c_{\nu r} \right] \\
\Delta_i =&\sum_{\alpha} \frac{\lambda_{\alpha i}^2}{\hbar \omega_{\alpha i}} \;\label{eq:delta2} .
\end{align}

$\tilde{H}'$ describes the perturbation to the exactly solvable Hamiltonian $\tilde{H}_0$. The 
perturbation consists of the hopping along the molecule and to and from the electrodes, where
the operators $\chi_i$ account for the creation and absorption of vibrations in the 
hopping processes. The perturbation can be considered small if either the hopping 
strengths $t_{ij}$ or $t^{r}$ are small and/or if the polaron binding energy $\Delta$ is large,
as the terms in a perturbative expansion are proportional to $1/\Delta$.

\subsection{Real-time expansion}
There are two limits to polaron transport, coherent band-like transport and incoherent
hopping transport. For weak electron-vibration coupling and low temperatures coherent
transport dominates, whereas for strong coupling and high temperatures transport is
a sequence of incoherent hopping processes. In this work we will focus on incoherent polaron 
hopping. To describe the physics in this regime we extend a formalism developed by 
B\"ottger and Bryksin\cite{Boettger85} for polaron transport in bulk systems 
to account for coupling to metallic electrodes.

To calculate quantities of interest, \eg the occupation number 
$\left\langle a_i^{\dagger}(t) a_i(t)\right\rangle $ and the current in a non-equilibrium
situation with applied bias, we make a real time expansion of the occupation number
along the Keldysh contour. The evolution in the interaction picture introduces the time 
dependence
\begin{align}
a_i(t)=&a_i e^{-i\left(\epsilon_{i}-\Delta_i\right)t}=a_i e^{-i\tilde{\epsilon}_{i}t}\nonumber\\
B_{i}(t)=&B_{i} e^{-i \omega_{i} t}.\nonumber
\end{align}
From here on we will use the shifted onsite energy 
$\tilde{\epsilon}_i=\epsilon_{i}-\Delta_i$ in all expressions.

The occupation number of the molecule can be written as
$\rho_{l}(t)=\left\langle a_l^{\dagger}(t) a_l(t)\right\rangle_{\tilde{H}}$. 
We express it in the interaction picture, assuming that the perturbation $\tilde{H}'$
is adiabatically turned on from the time $t_0=-\infty$, 
\begin{align}
\rho_{l}(t)=
\left\langle U^{\dagger}_{\tilde{H}_0}(t,-\infty)a_l^{\dagger}a_l U_{\tilde{H}_0}(t,-\infty) 
\right\rangle _{\tilde{H}_0} \nonumber
\end{align}
with time evolution operator 
\begin{align}
U_{\tilde{H}_0}(t,-\infty)=\mathrm{T}\left\lbrace \exp\left[-i\int_{-\infty}^t dt\tilde{H}'_{\tilde{H}_0}(t)
\right]\right\rbrace. \label{eq:timeevo}
\end{align}

\begin{figure}[htp]
 \centering
 \includegraphics[width=5cm]{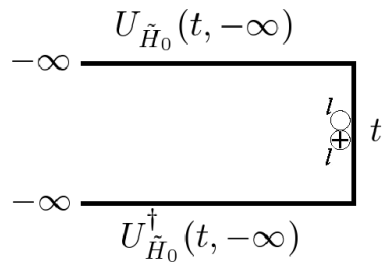}
 \caption{Schematic drawing of the Keldysh contour and the forward and backward time-evolution
operators. The open and crossed circle (the clamp) represent the two operators $a_l$ and $a_l^{\dagger}$, respectively,
which are evaluated at time $t$.\label{fig:keldysh}}
\end{figure}

A Taylor expansion of the time evolution operators in $\tilde{H}'$ defines a diagrammatic expansion. The 
forward time-evolution operator $U_{\tilde{H}_0}(t,-\infty)$ is expanded on the upper branch of 
the Keldysh contour, whereas the backward time-evolution operator $U_{\tilde{H}_0}^{\dagger}(t,-\infty)$ is 
expanded on the lower branch (see Fig.~\ref{fig:keldysh}). The index $\tilde{H}_0$ indicates that these 
operators are written in the interaction picture.
The time ordering operator `T' in Eq.~\ref{eq:timeevo} (anti-time ordering operator `$\tilde{\rm T}$')
ensures that the different times $t_i$, arising from the Taylor expansion of the forward (backward) time 
evolution operator, are ordered in the correct way along the contour. Note, oftentimes forward and 
backward time evolution operators are combined and a contour ordering operator `${\rm T_k}$' is 
introduced to ensure the correct ordering of times along the Keldysh contour.\cite{Jauho,Rammer86}

In performing the expansion in the time evolution operators, we obtain certain operator products, which
we have to average thermally. Since $\tilde{H}_0$ is quadratic in the fermion operators,
these can be treated using Wick's theorem. On the other hand, the vibrational operator products, involving 
various operators $\chi_i(t_j)$, cannot be factorized. The rules for the evaluation of these operator 
products are given in Appendix~\ref{app:op_prod}. 

\begin{figure}[thp]
 \centering
 \includegraphics[width=8cm]{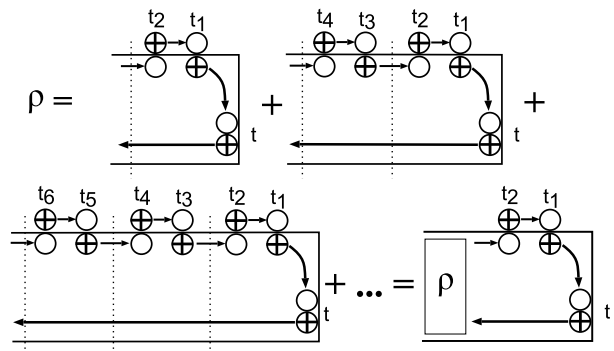}
 \caption{Ladder summation of a second order irreducible block representing two tunneling processes.
The irreducible blocks are separated by free sections (indicated by dotted lines). The summation can be
recast into a self-consistency equation for the occupation number $\rho$.\label{fig:bsp_rho_1}}
\end{figure}

A specific term in the Taylor expansion, 
is represented by a diagram with a certain number of vertices on the upper and lower branch
of the Keldysh contour, where each vertex is proportional either to $t_{ij}$ (a hopping 
vertex) or $t^{r}$ (a tunneling vertex). Each vertex consists of one open circle $\bigcirc$ 
(symbolizing a destruction operator) and one crossed circle $\bigoplus$ (symbolizing a creation 
operator). All circles belonging to operators acting on the molecule are drawn on the inside 
of the contour, whereas circles belonging to electrode operators are drawn on the outside of 
the contour (compare \eg Fig.~\ref{fig:bsp_rho_1}). The different vertices are connected by 
fermion (solid) and vibrational (dashed) lines and belong to different times $t_i$, which have 
to be (anti-) time ordered along the (lower) upper branch of the contour.

A feature of this expansion is that certain diagrams are diverging even in first order. These
diagrams can be identified by so called ``free sections'' (indicated by the dotted lines in 
Fig.~\ref{fig:bsp_rho_1}). A free section is a part of the diagram between two vertices 
(except for the clamp) where a vertical line can be drawn such that only internal fermion lines are cut
(neither a phonon line or an external fermion line belonging to the electrodes). 
In such a case, the vertical line always cuts an even number of internal fermion lines, 
as many left- as right-going. These left and right-going lines are pairwise associated with the 
same site. In the evaluation of such a diagram this leads to a divergence.\cite{note1}
Thus an infinite number of diagrams has to be summed up in a way similar to a 
`ladder'-approximation~\cite{Konstantinov61,Boettger85,Boettger93}. The regions in between free 
sections (excluding the clamp) are called irreducible blocks. They do not diverge.

In Figure~\ref{fig:bsp_rho_1} such a ladder-summation of a second order irreducible block
representing a tunneling process is shown. Similar to a Dyson series, the summation of
an infinite number of diagrams can be written as a self consistent equation for the occupation number
$\rho$.

For the interacting system we consider, certain diagrams 
lead to equations coupling the occupation number to many-particle correlation functions.
This is illustrated in Figure~\ref{fig:bsp}. The dotted vertical lines denoted
(i) and (ii) indicate free sections which lead to divergences. 
Let us concentrate on the free section (i) in Fig.~\ref{fig:bsp_rho}. 
To cure the divergence due to (i) a ladder summation has to be performed 
over all diagrams with the same divergence, \ie the same free section. 
This can be done in a complete and tractable manner by the introduction of many-particle correlation functions. 

Aside of this technical argument, there are also simple physical arguments why (and when)
the two-particle correlation functions affect the behavior of the occupation numbers (and the current).
Consider a hopping process from sites $m$ to site $n$: the hopping probability is determined by a second order irreducible diagram and the occupation of the initial and final site. Two-particle correlation functions express 
the probability to find e.g. the initial site $m$ occupied and the final site $n$ empty, such that the
hopping process can succeed. In general, the occupation of different sites is correlated, except in the trivial case when there is exactly one particle on the (central) system. If the charge density is finite, but very low, the charges are well described by a Boltzmann distribution\cite{Boettger75,Boettger76} and the two-particle correlation functions can be factorized in a ``Hartree-Fock'' type of approximation. This approach was taken in earlier works.\cite{Boettger75,Boettger76,Boettger93,Schmidt08}

\subsection{A hierarchy of many-particle correlation functions}
\begin{figure}
\centering
\subfigure[]{
    \label{fig:bsp_rho}
    \includegraphics[width=5.5cm]{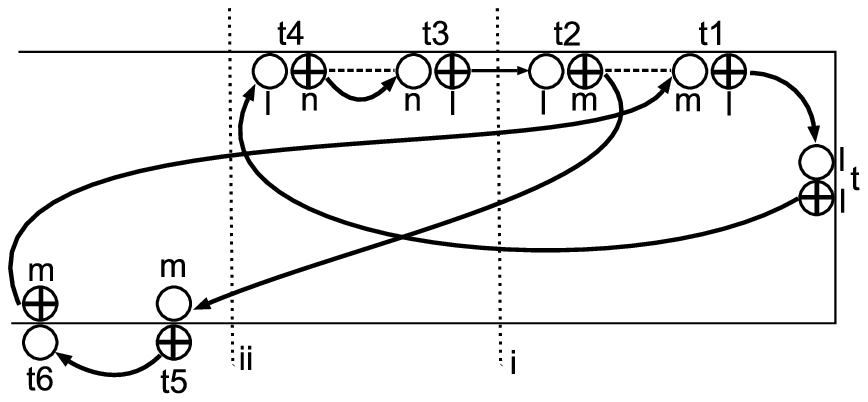}
  }\\
  \subfigure[]{
    \label{fig:bsp_drho}
    \includegraphics[width=5.5cm]{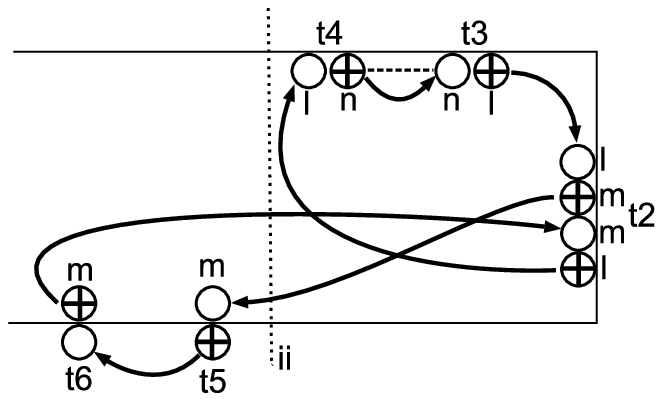}
  }
\caption{(a) Sixth order diagram in the real-time expansion of the occupation number
$\left\langle U_{\tilde{H}_0}^{\dagger}(t) a_{l}^{\dagger} a_{l} U_{\tilde{H}_0}(t) \right\rangle$
(b) Fourth order diagram in the real-time expansion of the two-particle correlation function $\left\langle U_{\tilde{H}_0}^{\dagger}(t_2) a_{l}^{\dagger}a_{m}a_{m}^{\dagger} a_{l} U_{\tilde{H}_0}(t_2) \right\rangle$. Free sections are indicated by the vertical dotted lines. \label{fig:bsp}}
\end{figure}

\begin{figure}
\centering
\includegraphics[width=5.5cm]{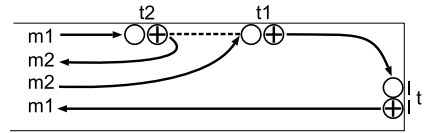}
\caption{A second order hopping diagram. The full lines represent fermion lines. The dashed line represents the sum of 
all possible vibrational lines arising from the diagrammatic rules, here it has 
a value $F_l^{+}(t_1-t_2)F_{m_2}^{+}(t_1-t_2)+F_l^{+}(t_1-t_2)+F_{m_2}^{+}(t_1-t_2)$.\label{fig:hop}}
\end{figure}

By inspection of the part of diagram Fig. 
\ref{fig:bsp_rho} {\it left} of the free section (i) one notices that this resembles a diagram  arising from
the real-time expansion of the two-particle correlation function $\left\langle U_{\tilde{H}_0}^{\dagger}(t_2) a_{l}^{\dagger}a_{m}a_{m}^{\dagger} a_{l} U_{\tilde{H}_0}(t_2) \right\rangle$, see Fig.~\ref{fig:bsp_drho}. Straightforward generalization shows that the ladder summation for the diagram~\ref{fig:bsp_rho} consists of all diagrams that arise from the real-time expansion of this particular two-particle correlation function, placed to the left of the free section (i), as indicated by Fig.~\ref{fig:hop}. 
In this way, an infinite number of diagrams to the occupation number $\rho$ can be accounted for 
by involving this particular two-particle correlation function with the particular irreducible block, given  
by the right part of the diagram Fig. \ref{fig:bsp_rho} from the free section indicated by (i) to the last vertex before the clamp. Other irreducible blocks involve other type of many-particle correlations
functions. What kind of correlation function is needed can be read off from the vertices of the irreducible block (following the contour) that are connected to the fermion lines crossing the free section (see also App.~\ref{app:corr_func}).  

Fig.~\ref{fig:bsp_drho} is only a particular diagram to the expansion of the two-particle correlation function $\left\langle U_{\tilde{H}_0}^{\dagger}(t_2) a_{l}^{\dagger}a_{m}a_{m}^{\dagger} a_{l} U_{\tilde{H}_0}(t_2) \right\rangle$. However, the arguments for replacing the free section (ii) in this diagram work the same way as for the expansion for the occupation number $\rho$. Therefore, we can write a linear equation for this two-particle correlation function involving all other two-particle correlation functions, the occupation number $\rho$ and other many-particle correlations functions. These many-particle correlations fulfill yet another set of linear equations. 

In this manner a hierarchy of linear equations is established. In principle, the approximation to the exact solution of the problem lies so far solely in the limited number of irreducible blocks that can be considered in a real calculation, i.e. in the order of expansion of the irreducible blocks in the hopping and and tunneling vertices. In practice, the question arises whether the hierarchy of equations can be solved exactly, or whether other approximations need to be applied, like a truncation or a factorization of the many-particle correlations functions.

\subsection{Truncation of the hierarchy}
\begin{figure}
\centering
\includegraphics[width=5.5cm]{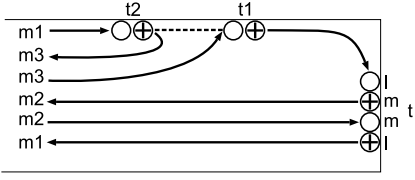}
\caption{A second order hopping diagram. The full lines represent fermions,
the dashed line represents the sum of all possible vibrational lines arising from the diagrammatic rules, 
here it has a value $F_m^{+}(t_1-t_2)F_{m_3}^{+}(t_1-t_2)+F_m^{+}(t_1-t_2)+F_{m_3}^{+}(t_1-t_2)$.\label{fig:dhop}}
\end{figure}

An important feature of the real-time expansion of the correlation functions is that 
certain diagrams vanish due to the Pauli exclusion principle. For example, the diagram 
depicted in Fig.~\ref{fig:dhop} arises in the expansion of the 
two-particle correlation function  and describes hopping process from site $m\equiv m_2$ to $m_3$ at time $t_2$ and back to $m$ at time $t_1$ . To 
its left the irreducible block is coupled to the three particle correlation function 
$\left\langle U_{\tilde{H}_0}^{\dagger}(0) a_{m_1}^{\dagger}a_{m_1} a_{m_2}^{\dagger} a_{m_3}a_{m_3}^{\dagger}a_{m_2}  U_{\tilde{H}_0}(0) \right\rangle$ with $m_1 \equiv l$.
For $m_3=m_1\equiv l$ there arises a special situation, the three particle correlation function 
is zero, since for fermions $\hat{n_l}(1-\hat{n_l})=0$ with $\hat{n}_l= a_l^{\dagger}a_l $. 

Furthermore, by similar arguments the Pauli exclusion principle leads to a {\it natural truncation} 
of the hierarchy of equations for {\it any finite} system. The $N$-particle correlation function 
$\left\langle U_{\tilde{H}_0}^{\dagger}(0) a_1^{\dagger}a_1a_2^{\dagger}a_2 \cdots a_N^{\dagger}a_N U_{\tilde{H}_0}(0)\right\rangle$ can not couple to any higher order correlation function
because they all vanish (recall that $N$ is the system size.)
Thus, a closed set of \emph{linear} equations for the real-time expansion of
all correlation functions can be constructed. The formal solution of this set of equations is simply 
a matter of back-substitution.

In the earlier works\cite{Boettger75,Boettger76,Boettger93,Schmidt08}, a Hartree-Fock factorization was applied in which products of particle number operators are replaced their expectation values. This leads to terms in the expansion that should not exist if the exact correlation functions were considered. 
The differences to the full theory are small, if the electron or hole densities are so low that they can 
be described by Boltzmann statistics.  However, there is the additional complication that the factorization 
leads to a {\it non-linear} self-consistency equation for the occupation numbers $\rho_i$. At finite bias 
it can be quite difficult to find a converging solution, especially as the system size becomes larger. 
This difficulty is avoided in our present theory where a solution to the linear equation set can be readily found by standard numerical methods. 

Summarizing the above, the relation 
\begin{align}
\frac{d}{dt}\rho_{l}(t)=&\int_{-\infty}^{t} dt_1\Bigg[ \sum_{m_1} 
\varrho_{m_1}(t_1) \mathcal{W}_{m_1l}(t_1-t)\nonumber\\
&+\sum_{m_1 m_2}\varrho_{m_1m_2}(t_1) \mathcal{W}_{m_1m_2l}(t_1-t)+\dots 
\Bigg]\label{eq:dd_density} 
\end{align}
for the time derivative of the occupation number is obtained.
The diagrammatic rules for construction and evaluation of irreducible blocks $\mathcal{W}$ are listed in appendix~\ref{app:diag_rules}. The generalized (one- and two-particle) correlation functions $\varrho_{m_1}$ ($\varrho_{m_1m_2}$) represent any order of the creation and destruction operators 
$a_{m_1},\,a_{m_1}^{\dagger}$ ($a_{m_1},\,a_{m_1}^{\dagger},\,a_{m_2},\,a_{m_2}^{\dagger}$) 
that arises in the free sections. Using the commutation relations for fermions, 
all generalized correlation functions $\varrho_{m_1 \ldots m_j}$ 
can be expressed by a sum of correlation functions $\rho_{m_1 \ldots m_i}$ (of the same or lower order) 
where we fix the  order of creation and destruction operators such that the creation operators at a site
are to the left of there destruction counterparts, i.e. $\rho_{m_1 \ldots m_i} \equiv a_{m_1}^{\dagger}a_{m_1} \ldots a_{m_i}^{\dagger}a_{m_i}$. In the rest of this article we 
only use these ordered correlation functions $\rho_{m_1 \ldots m_i}$ (note that the occupation number is 
naturally defined by the one-particle correlation function $\rho_{m_1}$).


\subsection{Explicit equations for the considered model}
In this article we want to calculate the stationary state of the system when a constant, finite bias is
applied between the metallic electrodes. Therefore, all correlation functions must be constant in time as
well, such that Eq.~\ref{eq:dd_density} reduces to
\begin{align}
0=\frac{d}{dt}\rho_{l}=& \sum_{m_1} 
\varrho_{m_1} \int_{-\infty}^{t} dt_1\mathcal{W}_{m_1l}(t_1)\nonumber\\
&+\sum_{m_1 m_2}\varrho_{m_1m_2} \int_{-\infty}^{t} dt_1\mathcal{W}_{m_1m_2l}(t_1)+\dots 
\label{eq:dd_steady} \; .
\end{align}
Furthermore, we restrict ourselves to the lowest non-vanishing order of diagrams, \ie
second order. Examples of two such diagrams in the expansion of the occupation
number and two particle correlation functions are given in Fig.~\ref{fig:dhop} and \ref{fig:hop}.
When comparing these two diagrams it is clear that they are mostly identical except for
two additional fermion lines in Fig.~\ref{fig:dhop} which enter as Kronecker-delta symbol. 
Of course, the corresponding irreducible blocks $\mathcal{W}$ couple to different correlation 
functions determined by the fermion lines leaving to and entering from the left. 

According to the `mirror' rule, to every diagram one can construct its complex conjugate by moving a vertex
from the upper part of the contour to the lower one (and vice versa). 
Following the rules given in the appendix the two diagrams (Fig.~\ref{fig:dhop} and \ref{fig:hop}) 
together with their respective complex conjugates and including the time integral of Eq.~\ref{eq:dd_steady} 
have the values
\begin{align}
\mathcal{W}_{m_1 m_2 l}=&-\frac{|t_{l,m_2}|^2}{\hbar^2}\mathcal{K}_{l}^2 \mathcal{K}_{m_2}^2 \int dt \, e^{\frac{i}{\hbar}\left(\tilde{\epsilon}_{l}-\tilde{\epsilon}_{m_2}\right)t}\nonumber\\
&\times \left[F_l^{+}(t)F_{m_2}^{+}(t)+F_l^{+}(t)+F_{m_2}^{+}(t)\right] \delta_{l m_1}
\end{align}
and
\begin{align}
\mathcal{W}_{m_1 m_2 m_3 l m}=-\frac{|t_{m,m_3}|^2}{\hbar^2} \mathcal{K}_{m}^2\mathcal{K}_{m_3}^2\int dt \, e^{\frac{i}{\hbar}\left(\tilde{\epsilon}_{m}-\tilde{\epsilon}_{m_3}\right)t}\nonumber\\
\times \left[F_m^{+}(t)F_{m_3}^{+}(t)+F_m^{+}(t)+F_{m_3}^{+}(t)\right] \delta_{l m_1}\delta_{m m_2}\,.
\end{align}
Explicit expressions for the constants $\mathcal{K}_{l}$ and the functions  $F_l^{+}(t)$ are given in the 
appendix \ref{app:diag_rules}.
The common element of all second order hopping diagrams in the expansion of the various 
correlation functions may be defined as
\begin{align}
\mathcal{W}_{l m}=&\frac{|t_{l,m}|^2}{\hbar^2}\mathcal{K}_{l}^2\mathcal{K}_{m}^2 \int dt \, e^{\frac{i}{\hbar}\left(\tilde{\epsilon}_{l}-\tilde{\epsilon}_{m}\right)t}\nonumber\\
&\times \left[F_l^{+}(t)F_{m}^{+}(t)+F_l^{+}(t)+F_{m}^{+}(t)\right]\,.
\end{align}
For the second order diagrams describing the hopping to and from the electrodes there exist also
such common elements. They are respectively
\begin{align}
W^{\rm L}_-=&\Gamma^{\rm L}  \int \frac{dE}{2\pi\hbar} (1-f_{\rm L}(E)) \mathcal{K}_{1}^2 \left(F_1^{+}(\tilde{\epsilon}_{1}-E)+\delta(\tilde{\epsilon}_{1}-E)\right)\nonumber \\
W^{\rm L}_+=&\Gamma^{\rm L} \int \frac{dE}{2\pi\hbar} f_{\rm L}(E) \mathcal{K}_{1}^2 \left(F_1^{+}(E-\tilde{\epsilon}_{1})+\delta(E-\tilde{\epsilon}_{1})\right) \label{eq:tunVL}\; ,
\end{align} 
where $\Gamma^{\rm L} = 2\pi |t^{\rm L}|^2 \rho_e$, 
$f_{\rm L}(E)$ is the Fermi function in left lead, and  $F_1^{+}(E)$ is the Fourier transform of $F_1^{+}(t)$.  
For the right interface a similar expression holds involving $f_{\rm R}(E)$ and $F_N^{+}(E)$.

Evaluating all second order diagrams simplified rate equations for the single-particle occupation number
can be stated
\begin{align}
\frac{d}{dt}\rho_{l}=& \sum_{m} \Big[ - \left(\rho_l-\rho_{lm}\right)\mathcal{W}_{l m}+ \left(\rho_m-\rho_{ml}\right)\mathcal{W}_{m l} \Big]\label{eq:rate_d}\\
\frac{d}{dt}\rho_{1}=& -\rho_{1} W^{\rm L}_{-}+\left(1-\rho_{1}\right) W^{\rm L}_{+}\nonumber\\
&-\left(\rho_1-\rho_{12}\right)\mathcal{W}_{1 2}+ \left(\rho_2-\rho_{21}\right)\mathcal{W}_{2 1} \; .\label{eq:rate_d2}
\end{align}
For the two-particle correlation functions on the inside of the molecule and connected to the left 
electrode the rate equations have similar form
\begin{align}
\frac{d}{dt} \rho_{lm} =\sum_n\bigg[ &-\left(\rho_{l m}-\rho_{l m n}\right)\mathcal{W}_{l n}+\left(\rho_{m n}-\rho_{l m n}\right)\mathcal{W}_{n l}\nonumber\\
&-\left(\rho_{l m}-\rho_{l m n}\right)\mathcal{W}_{m n}+\left(\rho_{l n}-\rho_{l m n}\right)\mathcal{W}_{n m}\bigg]\label{eq:rate_dd}
\end{align}
and
\begin{align}
\frac{d}{dt} \rho_{12} =&-\rho_{12} W^{\rm L}_{-}+ \left(\rho_{2}-\rho_{12}\right) W^{\rm L}_{+}\nonumber\\
&-\left(\rho_{12}-\rho_{123}\right)\mathcal{W}_{23}+\left(\rho_{13}-\rho_{123}\right)\mathcal{W}_{32}
\;.\label{eq:rate_dd2}
\end{align}
All other rate equations not presented above are constructed in an analogous way. 
This leads to a closed set of linear equations for all correlation functions up to the order of the system size $N$.

The theory is current conserving as Eq.~\ref{eq:rate_d} and \ref{eq:rate_d2} are the continuity equations
for the occupation of site $l$ and 1, respectively, which equal zero in the steady state situation we consider.
Thus the current can be computed at any point of the entire system. For convenience, we 
choose to compute the current through the left lead given by
\begin{align}
I_{\rm L}=e  \Big[& -\rho_{1} W^{\rm L}_- +\left(1-\rho_{1}\right) W^{\rm L}_+ \Big]\,.
\label{eq:curr}
\end{align}
Note that all many-particle correlation functions drop out of this expression. They influence the 
current only via their effect on the occupation at the first site, $\rho_{1}$.

\subsection{Polaron hopping transport in DNA}
In this section we apply the above presented diagrammatic approach to polaron 
hopping transport in DNA, which we have already studied earlier in a more simplified approach, 
involving the Hartree-Fock factorization.\cite{Schmidt08} DNA consists of a sequence of base pairs 
adenine and thymine and guanine and cytosine, which form a double helical ladder structure.
The electronic properties of DNA are determined by the HOMO (highest occupied 
molecular orbital), situated on the guanine and adenine bases, and the LUMO (lowest unoccupied 
molecular orbital), situated on cytosine and thymine bases.\cite{Endres04}

To describe polaron-hole hopping the relevant molecular orbital is the HOMO. We describe the
DNA chain in a minimal tight-binding model where each tight-binding site represents one 
HOMO either on a guanine or an adenine base. Both on-site energies $\epsilon_i$ and hopping 
integrals $t_{ij}$ depend on the base pair sequence.
For the direction-dependent hopping matrix elements $t_{ij}$ we use the values 
obtained from density functional theory by Siebbeles {\sl et al}.\cite{Senthilkumar05}
Adapting these values to our model of base pairs we obtain the next-neighbor hopping 
elements listed in table~\ref{tab:hopping}.
\begin{table}
5'-XY-3'(all in eV)\\
\begin{tabular}{|c|c|c|c|c|}\hline 
X$\diagdown$ Y & G  & C  & A  & T \\ \hline 
G & 0.119  & 0.046  & -0.186  & -0.048 \\ \hline 
C &  -0.075 & 0.119  & -0.037  & -0.013 \\ \hline 
A & -0.013 & -0.048  & -0.038  & 0.122 \\ \hline 
T & -0.037 & -0.186  & 0.148  & -0.038 \\ \hline 
\end{tabular}
\caption{Hopping integrals $t_{ij}$ taken from Ref.~\onlinecite{Senthilkumar05} 
and adapted to our model. The notation 5'-XY-3' indicates the direction
along the DNA strand (see, e.g., Fig.~1b in Ref.~\onlinecite{Endres04}.) \label{tab:hopping}}
\end{table}

As shown by Alexandre {\sl et al.}\cite{Alexandre03} and also other 
authors\cite{Conwell00,Henderson99,Joy06} polarons are formed on DNA molecules,
although the size of these polarons is still controversial. 
Fits to the temperature dependence of the linear conductance of experiments on long ($> 1000$ base pairs) DNA segments like Ref. \onlinecite{Yoo01} support the idea of small polaron formation with a local DNA distortion.~\cite{Triberis05,Triberis09} In this work we assume that
the size of polarons is restricted to single DNA base pairs. Such small polarons are formed due
to strong coupling of the electronic degrees of freedom to local vibrational modes of the DNA base pairs.
Exemplary, we consider only a single vibrational mode per base-pair, the so-called stretch 
modes with frequencies $\hbar \omega_i=16\,\rm{meV}$ for a GC base pair and 
$\hbar \omega_i=11\,\rm{meV}$ for an AT base pair.\cite{Starikov05}

The electron-vibration coupling strengths are chosen in such a way that the reorganization energy 
or polaron shifts (compare Eq.~\ref{eq:delta2}), $\Delta_{\rm A}=0.18\,\rm{eV}$ 
and $\Delta_{\rm G}=0.47\,\rm{eV}$, fit the values extracted from experiments and 
listed by Olofsson {\sl et al}.\cite{Olofsson01}.
These values probably underestimate the effect of the solvent on the reorganization energy.

To ensure energy dissipation and thermal occupation of the vibrations each base pair $i$ is 
coupled to a local environment, $H_{i,\rm bath}$, the microscopic details of which do not matter. 
This coupling changes the vibrations' spectra from discrete modes $\omega_i$ to continuous spectra,
\begin{align}
D_i(\omega)=&\frac{1}{\pi}\left(
  \frac{\eta_i(\omega)}{(\omega-\omega_i)^2+\eta_i(\omega)^2}-
\frac{\eta_i(\omega)}{(\omega+\omega_i)^2+\eta_i(\omega)^2}\right)\, ,
\end{align}
with frequency dependent broadening $\eta_i(\omega)$.~\cite{Galperin04} The actual form of 
$\eta_i(\omega)$ depends on the properties of the bath. A reasonable choice which assures also 
convergence at low and high frequencies is $\eta_i(\omega)=\eta_0 \, \frac{\omega^3}{\omega_i^3} \, \theta(\omega_{c}-\omega)$ with $\eta_0=0.5\,{\rm meV}$ and a cutoff of the order of 
$\hbar \omega_{c}=0.045\,{\rm eV}$.
To account for the spectral function of the base pair vibrations, the substitution
$\sum_{\alpha}\rightarrow \int d\omega D_i(\omega)$
has to be made in all equations introduced above.

\section{Results}
\label{sec:results}
The main focus of this article is the investigation of the influence of
correlations on the electronic transport characteristics. To do so, we compare our present theory with
exact correlation effects to our previous approach involving a Hartree-Fock factorization that deals
with correlations only on the mean-field level. For simplicity, we denote the latter approach as 
``mean-field correlations''.

It should be noted that correlation effects do not influence the transport properties 
of homogeneous DNA sequences or other homogeneous molecules.
This is because for a homogeneous system, not only $\rho_{lm} =  \rho_{ml}$ (this is always true by 
definition of the correlation functions), but also the hopping rates for forward and backward hopping 
processes are identical, $\mathcal{W}_{lm} =  \mathcal{W}_{ml}$. Therefore, the two-particle correlation 
functions drop out of Equations~\ref{eq:rate_d} and \ref{eq:rate_d2} for the occupation numbers, and 
consequently do not influence the current. The \iv curves are therefore identical, no matter whether 
exact correlations or mean-field correlations are considered.

\subsection{Zero bias conductance}
\label{subsec:zb_cond}
\begin{figure}
\includegraphics[width=8cm]{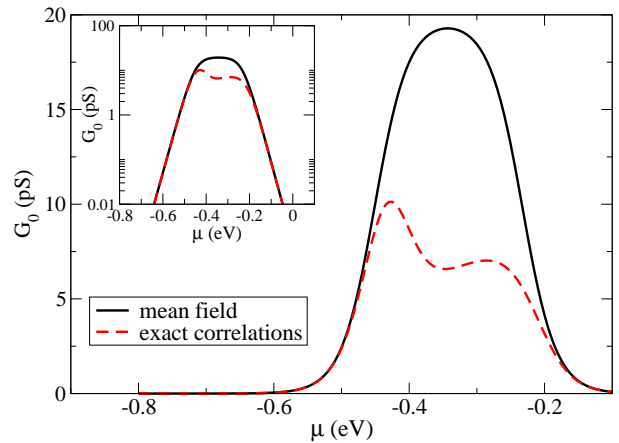}
\caption{Zero bias conductance $G_0$ as a function of chemical potential $\mu$ for the DNA 
molecule with sequence AAAGAAAA with mean-field correlations (black solid line) 
and exact correlations (red dashed line). The parameters used are 
$\epsilon_{\rm A}=-0.26\,\rm{eV}$, $\epsilon_{\rm G}=+0.25\,\rm{eV}$ relative to the
zero point of the chemical potential, polaron shifts 
$\Delta_{\rm A}=0.18\,\rm{eV}$ and $\Delta_{\rm G}=0.47\,\rm{eV}$,
symmetric coupling to leads with linewidths $\Gamma_{\rm L}=\Gamma_{\rm R}=0.001\,\rm{eV}$,
vibrational energies $\hbar \omega_{\rm A}=11\,\rm{meV}$, $\hbar \omega_{\rm G}=16\,\rm{meV}$, and
room temperature $k_{\rm B}T=25\,\rm{meV}$. The inset shows the same plot on logarithmic scale to stress that both models agree for very low and high occupation.\label{fig:AAAGAAAA}}
\end{figure}

As noted earlier the difference between the two approaches is small, when the occupation of holes or electrons is small, so that the occupation is described by a Boltzmann distribution function. 
To demonstrate this Fig.~\ref{fig:AAAGAAAA} shows the zero bias conductance $G_0$ as a function of chemical
potential $\mu$ for a DNA molecule with sequence AAAGAAAA with mean-field correlations 
(black solid line) and exact correlations (red dashed line). We have chosen of $\mu = 0 $ 
to lie above the (polaron shifted) guanine and adenine states, \ie in the HOMO-LUMO gap of DNA.
As can be seen from the Figure, especially from the inset on logarithmic scale, 
for values of $\mu<-0.5\,\rm{eV}$ and above $\mu>-0.15\,\rm{eV}$  the two curves agree with each other. 
In the first region the electron occupation number is very small, whereas in the latter the hole
occupation is very low. On the other hand, in the central region of the plot both curves differ strongly. The zero bias conductance is much lower when correlation effects are exactly accounted for. 
Furthermore, the red curve exhibits two maxima around the chemical potentials that 
agree with the onsite energies of adenine 
($\tilde{\epsilon}_{\rm A}$) and guanine ($\tilde{\epsilon}_{\rm G}$). 
The black curve, with mean-field correlation effects, only shows a single broad maximum. 
Similar to coherent quantum transport through molecules, the mean-field type approximation of correlations
overestimates the (zero bias) conductance and current and only shows a very simplified energetical structure.

\begin{figure}
\includegraphics[width=8cm]{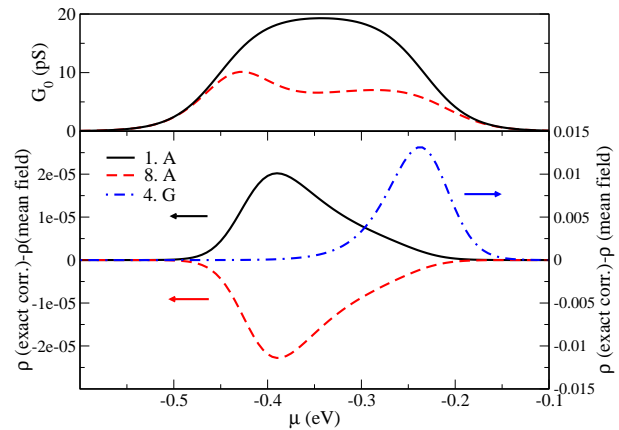}
\caption{Lower graph: Difference of the occupation for exact and mean-field correlations for 
the first (black solid line) fourth (red dashed line) and eighth 
(blue dash-dotted line) base of the DNA molecule AAAGAAAA as a function of chemical potential $\mu$. 
The used transport bias is $V_{\rm b}=0.01\,\rm{V}$, which is in the linear regime. As reference 
in the upper graph the plot of Fig.~\ref{fig:AAAGAAAA} is repeated. Parameters as in 
Fig.~\ref{fig:AAAGAAAA}.\label{fig:AAAGAAAA_rho}} 
\end{figure}

As an illustration, the effect of exact correlations on the occupation of the bases for the
sequence AAAGAAAA is shown in Fig.~\ref{fig:AAAGAAAA_rho}. The curves show the difference 
of the occupation for the two approaches for the first (black solid line) fourth 
(red dashed line) and eighth (blue dash-dotted line) base of the DNA molecule AAAGAAAA as a function of 
chemical potential $\mu$. The values were calculated for a small transport bias $V_{\rm b}=0.01\,\rm{V}$, 
which is in the linear regime. The sign and magnitude of the occupation difference changes with 
the bias direction and strength, respectively.

The maximum in occupation difference for the adenine and guanine bases is reached for chemical potential 
values which are close to the onsite energies of either adenine or guanine, respectively. The effect of the
correlation is the strongest for the isolated guanine base in the center of the sequence, where
the maximum difference in occupation is three orders of magnitude greater, than for the adenine bases. The 
occupations of the other adenine bases (not shown), are similar to the ones depicted in 
Fig.~\ref{fig:AAAGAAAA_rho}.

\subsection{Finite bias differential conductance}
\label{subsec:diff_cond}
In experiments a variation of the chemical potential is rather difficult to achieve. In molecular  electronics this is usually achieved by using a back gate electrode under an insulating substrate. 
However, for DNA the complication arises that its conformational structure is much influenced by 
the surface potential of the substrate and the electric potentials of the back gate. 
Oftentimes, though,  the \iv characteristics can be probed in setups where the molecules are in free suspension (mechanical break junctions) or standing upright in molecular monolayers. 

\begin{figure}
\includegraphics[width=8cm]{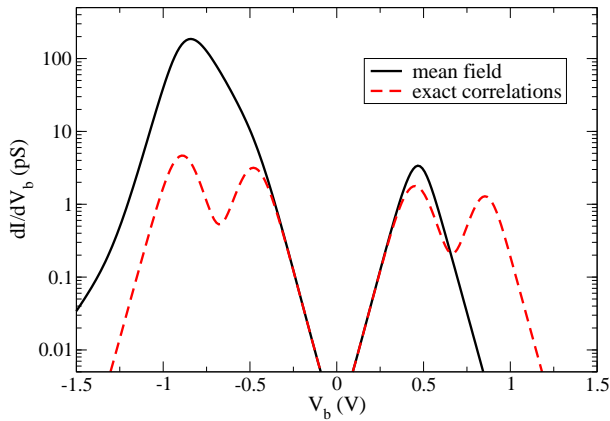}
\caption{Differential conductance $dI/dV_{\rm b}$ (logarithmic scale) as a function of applied bias
$V_{\rm b}$ for the DNA molecule with sequence AAAGAAAA, for mean-field correlations (black solid line) 
and exact correlations (red dashed line). Parameters as in Fig.~\ref{fig:AAAGAAAA}, 
but with the equilibrium chemical potential of the electrodes fixed at $\mu =0\,\mathrm{eV}$.
\label{fig:AAAGAAAA_iv}}
\end{figure}

When applying a transport bias over the molecule, the
occupation of the various molecular segments (base pairs in the case of the DNA) depends on the
magnitude of the applied bias ``felt'' at the particular location. The corresponding potential profile
can be interpreted as a ``local chemical potential'' that changes with the applied bias. As this local 
chemical potential moves over the energies of the base pair levels, similar effects as in 
Fig.~\ref{fig:AAAGAAAA} are expected. 
In Fig.~\ref{fig:AAAGAAAA_iv} we show the differential conductance 
$dI/dV_{\rm b}$ as a function of applied transport bias $V_{\rm b}$. The red curve including
exact correlation effects shows two maxima for both positive and negative bias, whereas the black 
line with mean-field correlations only has single peaks. For small bias both curves agree very well, as the
charge carrier occupation in this regime is very low. Note that again the mean-field approach overestimates
the (differential) conductance everywhere, except for large positive bias, where the second peak exists
in a bias region where the mean-field approach shows exponentially small conductance.

From these calculations it becomes clear that at finite transport bias the charge carrier occupations 
have value regions, where mean-field type of correlations are insufficient to describe polaron hopping 
transport through molecules like DNA.

We now discuss the position of the maxima in the zero bias and differential
conductance. For a homogeneous molecule, both $G_0$ and $dI/dV_{\rm b}$ only have a single
maximum, namely when the chemical potential (or the transport bias) is in resonance
with the level energy of the DNA base pairs ($\tilde{\epsilon}_{\rm G}=-0.22\,\rm{eV}$ and 
$\tilde{\epsilon}_{\rm A}=-0.44\,\rm{eV}$). 
Including exact correlations, for the DNA molecule AAAGAAAA there are two maxima
in the zero bias conductance as a function of chemical potential, or in the differential
conductance, for both positive and negative transport bias (see Figure~\ref{fig:AAAGAAAA} 
and \ref{fig:AAAGAAAA_iv}). These maxima can also be associated with the level energies 
of the two different types of bases, adenine and guanine. However, the exact positions of the
maxima in the zero bias conductance deviate slightly from their expected resonance 
positions due to charge rearrangement effects between guanine and adenine bases. Since the
hopping rates increase strongly with temperature, the charge rearrangement also increases.
Thus the position of the maxima is temperature dependent.

For the differential conductance the positions of the maxima
are shifted more strongly as a finite transport bias will lead to much stronger charge
rearrangements (``polarization''). This effect is increasingly important for the maxima at 
higher bias, \ie the second maxima are shifted more strongly from the anticipated resonance 
positions of the adenine bases energy as compared to the first maxima relating to the guanine 
energies. The charge rearrangement and therefore the position of the conductance maxima is 
quite sensitive to the considered DNA sequence (see also discussion in Ref.~\onlinecite{Schmidt08}).

\subsection{Sequence effects}
\label{subsec:sequence_effects}
\begin{figure}
\includegraphics[width=8cm]{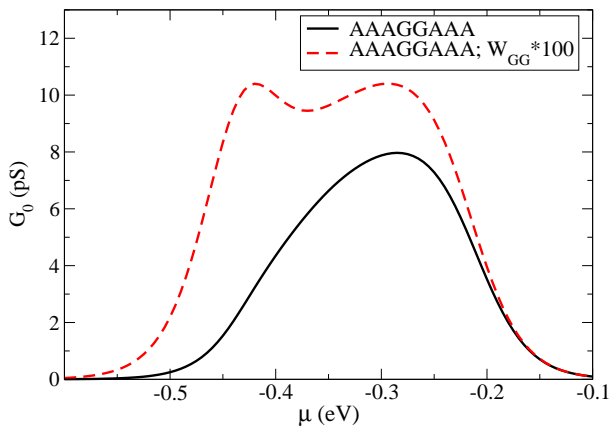}
\caption{The zero bias conductance $G_0$ as a function of chemical potential $\mu$ for the DNA 
sequence AAAGGAAA (black solid line) with exact correlations. Only a single maximum around the guanine 
energy is displayed. For the dashed line the hopping rate between the two central guanine bases is enhanced 
by a factor 100. This shows that the maximum at the adenine energy in the sequence AAAGGAAA 
is suppressed due to the peculiar hopping parameters relevant for DNA. Other parameters as in Fig.~\ref{fig:AAAGAAAA}. \label{fig:AAAGAAAA_vergl}}
\end{figure}

Are there always as many maxima as different species of molecular segments with different 
onsite energies, if exact correlations are considered?  The black solid curve in 
Figure~\ref{fig:AAAGAAAA_vergl} shows the zero bias conductance as a function of 
chemical potential for the DNA sequence AAAGGAAA.
This sequence is only a slight modification as compared to AAAGAAAA studied above, 
nevertheless the zero bias conductance exhibits only a single maximum, the `adenine'-peak is 
missing. A similar behavior is observed for the differential conductance as a function of 
applied bias.

Does this mean that for this sequence the adenine resonance does not occur? It turns out that 
it is just suppressed. The red dashed curve in Fig.~\ref{fig:AAAGAAAA_vergl} 
is calculated for the same sequence, but with a hopping rate 
$\mathcal{W}_{\rm GG}$ enhanced by a factor 100. In this curve again two maxima are visible, both in the
zero bias and differential conductance. Obviously, the adenine resonances were present 
but the resulting transport is strongly suppressed by the low hopping rate $\mathcal{W}_{\rm GG}$ 
for our model of DNA parameters.

At first glance, it might appear strange that the hopping rate between two guanine base pairs is 
so much limiting the transport at energies related to the adenine base pairs. 
However, this effect can be easily explained by looking at the current between the two guanine bases 
of the above DNA sequence. From Eq.~\ref{eq:rate_d} one can deduce
\begin{align}
{I}=&\, e \left(\rho_4-\rho_{45}\right)\mathcal{W}_{45}+ e \left(\rho_5-\rho_{45}\right)\mathcal{W}_{54} \; \nonumber\\
=&\, e \left(\rho_4-\rho_5\right)\mathcal{W}_{\rm GG}\label{eq:curr_rho}\;,
\end{align}
where the sites $4$ and $5$ denote the two guanine bases and for the second line 
$\mathcal{W}_{45}=\mathcal{W}_{54}=\mathcal{W}_{\rm GG}$ was used. 

As the current is conserved, the above equation is identical to the current obtained from 
Eq.~\ref{eq:curr}. Obviously, a small rate $\mathcal{W}_{\rm GG}$ leads in general to
a small current. However, the important matter lies in the difference of occupations
$(\rho_4-\rho_5)$ of the two guanine bases (for a small applied transport bias) that varies 
strongly depending on the chemical potential. For the DNA sequence AAAGGAAA the current 
reaches its maximum value when the difference in occupation between the two guanine sites is 
the greatest. This is the case when the guanine onsite energy is in resonance with the chemical 
potential, \ie around $\mu=-0.22\,\rm{eV}$, as at this energy the guanine occupation changes 
rapidly from unity to zero while lowering the chemical potential. For much lower values of the 
chemical potential (in particular $\mu\approx-0.44\,\rm{eV}$) the occupation of the guanine bases 
is already close to zero 
and thus the occupation {\it differences} are also small. The double guanine segment limits 
the current the molecule can support at the adenine energy around $-0.44\,\rm{eV}$. 

If we artificially enhance the rate $\mathcal{W}_{\rm GG}$ by a factor 100, see dashed line 
on Fig.~\ref{fig:AAAGAAAA_vergl} the conductance is overall increased, though not by a factor 
of 100. The increase is much stronger around the adenine energy than around the guanine energy. 
Removing the original bottleneck, even small occupation differences between the guanines lead now 
to a fairly sized conductance around the adenine energy. As the conductance at the guanine energy 
increases only slightly {\it despite} the 100-fold increased rate, the guanine occupation difference 
at the guanine energy is actually {\it reduced} almost by the same factor.  

Similar argument hold again for the suppression of the adenine related maxima in 
the differential conductance.
%

\section{Summary}
\label{sec:summary}

We have presented a diagrammatic real-time approach to polaron hopping through molecules coupled to metallic electrodes, taking into account vibration-mediated charge correlations. This technique leads to a hierarchy
of linear rate equations for the occupation and many-particle correlation functions, which
is naturally truncated for a finite size system. Thus, an exact description of correlation effects is
possible for a given order of the perturbation expansion in the hopping parameters. Using short DNA molecules as an example, we show that including exact correlations lowers the zero bias conductance of inhomogeneous DNA sequences when the average charge occupation is sufficiently high. For  exponentially small charge occupation, a mean-field description of the correlations is adequate. For the 
\iv characteristics, the inclusion of exact correlation effects is 
necessary since at the experimentally relevant bias voltage the local 
charge densities are generally non-negligible. 
We further have shown that for short DNA molecules consisting of two different types of base pairs, the zero bias and differential conductance shows two maxima. Depending on the specific
sequence, one of these maxima can be suppressed due to low hopping rates.

{\em Acknowledgments.} We thank the  \mbox{Landesstiftung} Baden-W\"urttemberg for
financial support via the \mbox{Kompetenznetz} ``Funktionelle
Nanostrukturen''.

\begin{appendix}
\section{}
\label{app:diag_rules}
\subsection*{Construction of irreducible block diagrams}
\label{sec:construction}
Below we will state the rules for the construction and evaluation of irreducible block diagrams.
The rules for pure hopping diagrams, \ie those containing only vertices $\propto t_{ij}$, were 
developed by B\"ottger and Bryksin \cite{Boettger85}. We extended their theory adding new rules 
to treat diagrams with tunneling vertices $\propto t_{i\nu}^{r}$.

The perturbative expansion can be visualized by the construction of diagrams which are equivalent
to expressions in the analytic expansion. The main contribution to the diagrams comes
from so called irreducible blocks, which, as the name implies, cannot be decomposed into more simple
diagrams. The main feature of an irreducible block diagram is, that it does not diverge, when 
integrating over the internal times $t_i$. Irreducible blocks can be identified by their property
of not allowing free sections. A free section is a vertical line drawn between the leftmost 
vertex and the rightmost vertex (except for the clamp) that does not cross either a 
phonon line or an external fermion (tunneling) line.

The rules come in two sets: the first for the construction and labeling of possible
diagram, the second set for the evaluation of a particular diagram. The rules are general 
for all orders of perturbation theory.
\begin{enumerate}
\item Draw the Keldysh time contour as a rectangle which is open to the left, corresponding to $t \rightarrow -\infty$.

\item For a diagram of order $n$ we draw on the contour $n+1$ pair vertices
consisting of one open circle $\bigcirc$ (symbolizing a destruction operator) and one 
crossed circle $\bigoplus$ (symbolizing a creation operator). 
All circles belonging to operators acting on the molecule are drawn on the inside of the contour, 
whereas circles belonging to electrode operators are drawn on the outside of the contour.
Therefore, if the pair vertex is due to a tunneling process $t^r$ one circle is on the inside
and the other one is on the outside of the contour. The circles of a hopping process are both drawn on 
the inside of the contour where the open circle is always `earlier' along the Keldysh contour than 
the crossed circle. As we calculate diagrams to evaluate the density matrix, we draw one pair vertex 
(also called `clamp'~\cite{Boettger85}) at the inside of the right vertical line of the Keldysh contour, 
corresponding to time $t$. The other $n$ vertices are drawn at $n$ times $t_i$ on either the upper or 
lower branch of the Keldysh contour (where $t_2$ is the leftmost, earliest time and $t_1$ is the rightmost,
latest time).

\item Each open circles $\bigcirc$ on the inside of the contour has one ingoing fermion line
(arrow pointing to the vertex) and each crossed circle $\bigoplus$ has one outgoing fermion line
(arrow pointing away from the vertex) which is locally directed along the Keldysh contour.

\item Complementary circles outside the contour are pairwise connected by a fermion line drawn 
{\it outside} of the contour. Since this line corresponds to an electron propagating in electrode 
$r$ the connected circles have to belong to the same electrode $r$, otherwise the diagram contribution 
is zero.
 
\item The clamp is always connected by a fermion line to the rightmost 
vertex (other than the clamp) drawn { \it inside} of the contour, \ie the
clamp and the rightmost vertex are associated with the same state.
If the rightmost vertex is a hopping vertex, the fermion line is directed 
along the contour. If the rightmost vertex is a tunneling vertex, the inside circle (open or crossed) 
is connected to the complementary circle of the clamp, no matter what the direction of the fermion line.

\item The remaining unconnected inside circles have fermion lines going into (coming from) 
the region left of the diagram ($t\rightarrow -\infty$) without intersecting each other. 

\item Each circle belongs to one specific state for the molecule or the electrode. 
We label the molecule states (sites) by latin characters (e.g. $m,n, \ldots$) and 
the electrode states by Greek characters (e.g. $\nu$). Note that the two circles of 
a hopping vertex can not correspond to the same state (site). Since we want to calculate the
density matrix $\rho_l$ then both circles of the clamp are associated with the
state (site) $l$. 

\item Except for the clamp, the circles on the inside of the contour must be connected 
by phonon lines so that the diagram has no free section, as defined above. One circle can be connected to more than
one phonon line. All diagrams with different number of phonon lines (but still without free sections)
have to be considered. Only circles belonging to the same state (site) can be connected by a phonon line.
Therefore, the two circles of a hopping vertex can not be connected.

\end{enumerate}

The rules for evaluating a diagram are as follows.

\begin{enumerate}
 \item A hopping vertex at time $t_i$ is
associated with a factor 
$\pm i t_{n m}\mathcal{K}_{n}\mathcal{K}_{m} e^{-i\left(\epsilon_{n}-\epsilon_{m }\right)(t_i-t_2)}$
where the creation operator (crossed circle) corresponds to site label $n$ and the destruction
operator to the site label $m$ (recall that $t_2$ is the leftmost time of the diagram).
A tunneling vertex is associated with a factor 
$\pm i t^{r}\mathcal{K}_{m} e^{-i\left(\epsilon_{\nu}-\epsilon_{m}\right)(t_i-t_2)}$ or 
$\pm i t^r\mathcal{K}_{n} e^{-i\left(\epsilon_{n}-\epsilon_{\nu}\right)(t_i-t_2)}$ if 
the creation operator acts on the electrode or on the molecule, respectively.
Vertices on the upper half of the contour have the minus sign, vertices on the 
lower half of the contour have the plus sign. 
The factor $$\mathcal{K}_{m}=\exp \left\lbrace -\frac{1}{2}\sum_{\alpha}
 \left(\frac{\lambda_{\alpha m}}{\omega_{\alpha}}\right)^2  \left(2 N(\omega_{\alpha})+1 \right) \right\rbrace.$$

\item The outside fermion lines of the electrodes $r$ contribute a factor $1-f^r_{\nu}$ or $f^r_{\nu}$ 
depending whether they run in the direction of the contour or against it. Here  $f^r_{\nu}$ is the 
Fermi function at energy $\epsilon_{\nu} -\mu_r$, with the chemical potential $\mu_r$.

\item The fermion lines entering (leaving) the irreducible
block from (to) the left are labeled from top to bottom, with one ingoing and
one outgoing line belonging to pairwise the same state. The labels determine the 
indices of the irreducible block, \eg  $\mathcal{W}_{m_1 m_2 l}(t)$. The line attached to
the clamp and leaving to the left is associated with a Kronecker delta function of the
states the line connects.

A phonon line connecting two circles both associated to a state (site) $m$ has a value 
\begin{align}
&F_m^{\zeta}(t_i-t_j)=\exp\left\lbrace \zeta A_m(t_i-t_j)\right\rbrace-1\; ,\mbox{with}\nonumber\\
&A_m(t)=\sum_{\alpha} \left(\frac{\lambda_{\alpha m}}{\omega_{\alpha}} \right)^2 \frac{\cos\left(\omega_{\alpha} \left[t+i\hbar\beta/2 \right]\right)}{\sinh\left(\hbar \omega_{\alpha}\beta/2 \right)}\;,\nonumber
\end{align}
where the circle at time $t_i$ is later on the contour than the circle at time $t_j$.
The factor $\zeta$ is determined by the type of circles the line connects.
If the circles are different $\zeta =+1$, otherwise $\zeta =-1$.

\item Multiply with a factor $(-1)^{M+N}$, where $M$ is the number of intersections of fermion lines 
on the \textit{outside} of the contour (tunneling lines) and $N$ is the number of intersections 
of fermion lines on the \textit{inside} of the contour.

\item We integrate over all internal times $t_i$ (except $t_1$ and $t_2$) and sum 
over all electrode states $\nu$ and all internal molecule states $i,j$, 
{\it except} the states associated with the clamp.
\end{enumerate}

\begin{figure}
\includegraphics[width=5.5cm]{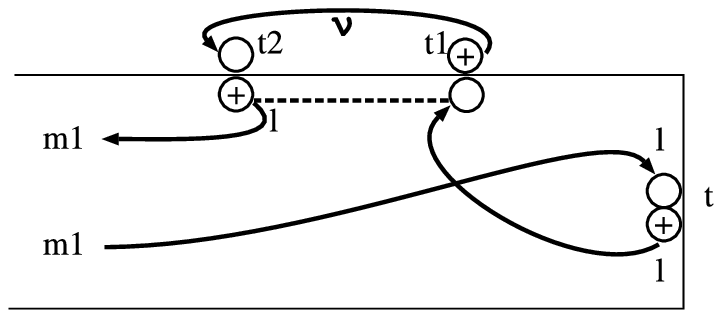}
\caption{\label{fig:ord2tun3}}
\end{figure}

As an example Fig.~\ref{fig:ord2tun3} shows a second order diagram for a hopping
process between molecule and metallic electrode.
Following the rules listed above the value of this diagram is
\begin{align*}
\mathcal{W}_{m_1,l}=&(-1)(-i)^2|t^r|^2 \delta_{m_1 l} \mathcal{K}_{l}^2\\ 
&\times \sum_{\nu} f_{\nu} e^{-i (\tilde{\epsilon_l}-\epsilon_{\nu})(t_1-t_2)}  F_l^{+}(t_1-t_2)\\
=& \Gamma^{r} \delta_{m_1 l} \mathcal{K}_{l}^2 \int \frac{dE}{2\pi\hbar} f_{r}(E)  F_l^{+}(\tilde{\epsilon}_{l}-E)\,.
\end{align*}
The last line is true in the wide band limit with $\Gamma^{\rm L,R}\propto \rho_e |t^{\rm L,R}|^2$.

\section{}
\subsection*{Determination of the correlation function}
\label{app:corr_func}
The correlation function with which the irreducible block is convoluted can be easily 
constructed from the diagrams. Lines leaving the block correspond to creation operators, 
lines entering the block correspond to destruction operators. The order of operators 
from left to right in the correlation functions corresponds to the order of the
lines leaving/entering the diagram from bottom to top.

For example, let us consider some irreducible block
$\mathcal{W}_{m_1m_2m_3l}(t)$ which has the following order of 
the terminals at the left of the diagram from bottom to top:
$m_3$ outgoing, $m_2$ ingoing, $m_3$ ingoing, $m_1$ outgoing, $m_2$ outgoing, 
and $m_1$ ingoing. Thus this rate will be convoluted with the third order correlation function 
\begin{align}
\left\langle U_{\tilde{H}_0}^{\dagger}(0) a_{m_3}^{\dagger} a_{m_2} a_{m_3}
a_{m_1}^{\dagger}a_{m_2}^{\dagger} a_{m_1}U_{\tilde{H}_0}(0)\right\rangle_{\tilde{H}}=\qquad\qquad\nonumber\\
\rho_{m_1 m_3}-\rho_{m_1 m_2 m_3}\,.\nonumber
\end{align}

\section{}
\subsection*{Vibrational operator products}
\label{app:op_prod}
In the perturbation expansion of the single particle density matrix $\rho_{l}(t)=\left\langle a_l^{\dagger}(t) a_l(t)\right\rangle_{\tilde{H}}$ to order $n$ in the perturbative Hamiltonian $\tilde{H'}$ (Eq.~\ref{eq:Hperturb2}),
one obtains up to $n$ vibrational operators (equal number of $\chi$ and $\chi^{\dagger}$) at different times 
which act upon the same vibrational states.
\be
\lefteqn{\left\langle \chi_k(t_1) \chi_{k}^{\dagger}(t_2)\chi_k(t_3)  \cdots  \chi_{k}^{\dagger}(t_{n}) \right\rangle_{H_0}=}\nonumber\\
& &\left(\exp \left\lbrace -\frac{1}{2}\sum_{\alpha} \left(\frac{\lambda_{k\alpha}}{\Omega_{k\alpha}}\right)^2 \left(2 N(\Omega_{k \alpha})+1 \right) \right\rbrace \right)^{n} \nonumber\\
& & \qquad\times \exp \bigg\lbrace  \left\lbrace \zeta_{12} A_k(t_1-t_2)\right\rbrace + \mathrm{T}_C \left\lbrace \zeta_{13} A_k(t_1-t_3)\right\rbrace  \nonumber\\
& & \qquad+\mathrm{T}_C \left\lbrace \zeta_{3,n} A_k(t_{3}-t_{n})\right\rbrace +\cdots \bigg\rbrace \nonumber\:,
\ee
where 
\begin{equation}
\zeta_{ij}=
\begin{cases}
+1 & \text{when } \chi_k(t_i) \chi_k^{\dagger}(t_j) \text{ or } \chi_k^{\dagger}(t_j) \chi_k(t_i),\\
-1 & \text{when } \chi_k(t_i) \chi_k(t_j) \text{ or } \chi_k^{\dagger}(t_i) \chi_k^{\dagger}(t_j)
\end{cases}\nonumber\;.
\end{equation}
The expression $\mathrm{T}_C$ in $\mathrm{T}_C \left\lbrace \zeta_{12} A_k(t_1-t_2)\right\rbrace$ ensures, that $t_1$ 
is later on the contour than $t_2$ and $A_k(t_1-t_2)$ is given by
\begin{align}
A_k(t_1-t_2)=& \sum_{\alpha}\left(\frac{\lambda_{k\alpha}}{\omega_{k\alpha}} \right)^2 \frac{\cos\left(\omega_{k\alpha} \left[t_1-t_2 +i\hbar\beta/2 \right]\right)}{\sinh\left(\hbar \omega_{k\alpha}\beta/2 \right)}  \nonumber\;.
\end{align}
For a correlator with $n$ operators $\chi_k$ and $\chi_k^{\dagger}$ acting on the same state
one gets $N=\frac{(n)(n-1)}{2}$ different terms $A_k(t_i-t_j)$ in the exponential function.
This is due to the various operator commutations involved in deriving the above expression.

\end{appendix}


\begin{thebibliography}{31}
\expandafter\ifx\csname natexlab\endcsname\relax\def\natexlab#1{#1}\fi
\expandafter\ifx\csname bibnamefont\endcsname\relax
  \def\bibnamefont#1{#1}\fi
\expandafter\ifx\csname bibfnamefont\endcsname\relax
  \def\bibfnamefont#1{#1}\fi
\expandafter\ifx\csname citenamefont\endcsname\relax
  \def\citenamefont#1{#1}\fi
\expandafter\ifx\csname url\endcsname\relax
  \def\url#1{\texttt{#1}}\fi
\expandafter\ifx\csname urlprefix\endcsname\relax\def\urlprefix{URL }\fi
\providecommand{\bibinfo}[2]{#2}
\providecommand{\eprint}[2][]{\url{#2}}

\bibitem[{\citenamefont{Yoo et~al.}(2001)\citenamefont{Yoo, Ha, Lee, Park, Kim,
  Kim, Lee, Kawai, and Choi}}]{Yoo01}
\bibinfo{author}{\bibfnamefont{K.~H.} \bibnamefont{Yoo}},
  \bibinfo{author}{\bibfnamefont{D.~H.} \bibnamefont{Ha}},
  \bibinfo{author}{\bibfnamefont{J.~O.} \bibnamefont{Lee}},
  \bibinfo{author}{\bibfnamefont{J.~W.} \bibnamefont{Park}},
  \bibinfo{author}{\bibfnamefont{J.}~\bibnamefont{Kim}},
  \bibinfo{author}{\bibfnamefont{J.~J.} \bibnamefont{Kim}},
  \bibinfo{author}{\bibfnamefont{H.~Y.} \bibnamefont{Lee}},
  \bibinfo{author}{\bibfnamefont{T.}~\bibnamefont{Kawai}}, \bibnamefont{and}
  \bibinfo{author}{\bibfnamefont{H.~Y.} \bibnamefont{Choi}},
  \bibinfo{journal}{Phys.\ Rev.\ Lett.} \textbf{\bibinfo{volume}{87}},
  \bibinfo{pages}{198102} (\bibinfo{year}{2001}).

\bibitem[{\citenamefont{Shigematsu et~al.}(2003)\citenamefont{Shigematsu,
  Shimotani, Manabe, Watanabe, and Shimizu}}]{Shigematsu03}
\bibinfo{author}{\bibfnamefont{T.}~\bibnamefont{Shigematsu}},
  \bibinfo{author}{\bibfnamefont{K.}~\bibnamefont{Shimotani}},
  \bibinfo{author}{\bibfnamefont{C.}~\bibnamefont{Manabe}},
  \bibinfo{author}{\bibfnamefont{H.}~\bibnamefont{Watanabe}}, \bibnamefont{and}
  \bibinfo{author}{\bibfnamefont{M.}~\bibnamefont{Shimizu}},
  \bibinfo{journal}{J.\ Chem.\ Phys.} \textbf{\bibinfo{volume}{118}},
  \bibinfo{pages}{4245} (\bibinfo{year}{2003}).

\bibitem[{\citenamefont{Choi et~al.}(2008)\citenamefont{Choi, Kim, and
  Frisbie}}]{Choi08}
\bibinfo{author}{\bibfnamefont{S.~H.} \bibnamefont{Choi}},
  \bibinfo{author}{\bibfnamefont{B.}~\bibnamefont{Kim}}, \bibnamefont{and}
  \bibinfo{author}{\bibfnamefont{C.~D.} \bibnamefont{Frisbie}},
  \bibinfo{journal}{Science} \textbf{\bibinfo{volume}{320}},
  \bibinfo{pages}{1482} (\bibinfo{year}{2008}).

\bibitem[{\citenamefont{Kubatkin et~al.}(2003)\citenamefont{Kubatkin, Danilov,
  Hjort, Cornil, Bredas, Stuhr-Hansen, Hedegard, and Bjornholm}}]{kubatkin03}
\bibinfo{author}{\bibfnamefont{S.}~\bibnamefont{Kubatkin}},
  \bibinfo{author}{\bibfnamefont{A.}~\bibnamefont{Danilov}},
  \bibinfo{author}{\bibfnamefont{M.}~\bibnamefont{Hjort}},
  \bibinfo{author}{\bibfnamefont{J.}~\bibnamefont{Cornil}},
  \bibinfo{author}{\bibfnamefont{J.}~\bibnamefont{Br\'edas}},
  \bibinfo{author}{\bibfnamefont{N.}~\bibnamefont{Stuhr-Hansen}},
  \bibinfo{author}{\bibfnamefont{P.}~\bibnamefont{Hedeg\aa{}rd}}, \bibnamefont{and}
  \bibinfo{author}{\bibfnamefont{T.}~\bibnamefont{Bj\o{}rnholm}},
  \bibinfo{journal}{Nature} \textbf{\bibinfo{volume}{425}},
  \bibinfo{pages}{698} (\bibinfo{year}{2003}).

\bibitem[{\citenamefont{Osorio et~al.}(2007)\citenamefont{Osorio, ONeill,
  Wegewijs, Stuhr-Hansen, Paaske, Bjornholm, and van~der Zant}}]{osorio07}
\bibinfo{author}{\bibfnamefont{E.~A.} \bibnamefont{Osorio}},
  \bibinfo{author}{\bibfnamefont{K.}~\bibnamefont{O'Neill}},
  \bibinfo{author}{\bibfnamefont{M.}~\bibnamefont{Wegewijs}},
  \bibinfo{author}{\bibfnamefont{N.}~\bibnamefont{Stuhr-Hansen}},
  \bibinfo{author}{\bibfnamefont{J.}~\bibnamefont{Paaske}},
  \bibinfo{author}{\bibfnamefont{T.}~\bibnamefont{Bj\o{}rnholm}},
  \bibnamefont{and} \bibinfo{author}{\bibfnamefont{H.~S.~J.}
  \bibnamefont{van~der Zant}}, \bibinfo{journal}{Nano Lett.}
  \textbf{\bibinfo{volume}{7}}, \bibinfo{pages}{3336} (\bibinfo{year}{2007}).

\bibitem[{\citenamefont{Galperin et~al.}(2004)\citenamefont{Galperin, Ratner,
  and Nitzan}}]{Galperin04}
\bibinfo{author}{\bibfnamefont{M.}~\bibnamefont{Galperin}},
  \bibinfo{author}{\bibfnamefont{M.~A.} \bibnamefont{Ratner}},
  \bibnamefont{and} \bibinfo{author}{\bibfnamefont{A.}~\bibnamefont{Nitzan}},
  \bibinfo{journal}{J.\ Chem.\ Phys.} \textbf{\bibinfo{volume}{121}},
  \bibinfo{pages}{11965} (\bibinfo{year}{2004}).

\bibitem[{\citenamefont{Alexandrov and Bratkovsky}(2007)}]{Alexandrov07}
\bibinfo{author}{\bibfnamefont{A.~S.} \bibnamefont{Alexandrov}}
  \bibnamefont{and} \bibinfo{author}{\bibfnamefont{A.~M.}
  \bibnamefont{Bratkovsky}}, \bibinfo{journal}{J.\ Phys.: Condens.\ Matter}
  \textbf{\bibinfo{volume}{19}}, \bibinfo{pages}{255203}
  (\bibinfo{year}{2007}).

\bibitem[{\citenamefont{Alexandrov and Bratkovsky}(2006)}]{Alexandrov06}
\bibinfo{author}{\bibfnamefont{A.~S.} \bibnamefont{Alexandrov}}
  \bibnamefont{and} \bibinfo{author}{\bibfnamefont{A.~M.}
  \bibnamefont{Bratkovsky}}, \bibinfo{journal}{cond-mat} p.
  \bibinfo{pages}{0603467} (\bibinfo{year}{2006}).

\bibitem[{\citenamefont{Mitra et~al.}(2005)\citenamefont{Mitra, Aleiner, and
  Millis}}]{Mitra05}
\bibinfo{author}{\bibfnamefont{A.}~\bibnamefont{Mitra}},
  \bibinfo{author}{\bibfnamefont{I.}~\bibnamefont{Aleiner}}, \bibnamefont{and}
  \bibinfo{author}{\bibfnamefont{A.~J.} \bibnamefont{Millis}},
  \bibinfo{journal}{Phys.\ Rev.\ Lett.} \textbf{\bibinfo{volume}{94}},
  \bibinfo{pages}{076404} (\bibinfo{year}{2005}).

\bibitem[{\citenamefont{Berlin et~al.}(2001)\citenamefont{Berlin, Burin, and
  Ratner}}]{Berlin01}
\bibinfo{author}{\bibfnamefont{Y.~A.} \bibnamefont{Berlin}},
  \bibinfo{author}{\bibfnamefont{A.~L.} \bibnamefont{Burin}}, \bibnamefont{and}
  \bibinfo{author}{\bibfnamefont{M.~A.} \bibnamefont{Ratner}},
  \bibinfo{journal}{J.\ Am.\ Chem.\ Soc.} \textbf{\bibinfo{volume}{123}},
  \bibinfo{pages}{260} (\bibinfo{year}{2001}).

\bibitem[{\citenamefont{Schmidt et~al.}(2008)\citenamefont{Schmidt, Hettler,
  and Sch\"on}}]{Schmidt08}
\bibinfo{author}{\bibfnamefont{B.~B.} \bibnamefont{Schmidt}},
  \bibinfo{author}{\bibfnamefont{M.~H.} \bibnamefont{Hettler}},
  \bibnamefont{and} \bibinfo{author}{\bibfnamefont{G.}~\bibnamefont{Sch\"on}},
  \bibinfo{journal}{Phys.\ Rev.\ B} \textbf{\bibinfo{volume}{77}},
  \bibinfo{pages}{165337} (\bibinfo{year}{2008}).

\bibitem[{\citenamefont{B\"ottger and Bryksin}(1985)}]{Boettger85}
\bibinfo{author}{\bibfnamefont{H.}~\bibnamefont{B\"ottger}} \bibnamefont{and}
  \bibinfo{author}{\bibfnamefont{V.~V.} \bibnamefont{Bryksin}},
  \emph{\bibinfo{title}{Hopping conduction in Solids}}
  (\bibinfo{publisher}{Akademie Verlag Berlin}, \bibinfo{year}{1985}).

\bibitem[{\citenamefont{Park et~al.}(2000)\citenamefont{Park, Park, Lim,
  Anderson, Alivisatos, and McEuen}}]{park00}
\bibinfo{author}{\bibfnamefont{H.}~\bibnamefont{Park}},
  \bibinfo{author}{\bibfnamefont{J.}~\bibnamefont{Park}},
  \bibinfo{author}{\bibfnamefont{A.~K.~L.} \bibnamefont{Lim}},
  \bibinfo{author}{\bibfnamefont{E.~H.} \bibnamefont{Anderson}},
  \bibinfo{author}{\bibfnamefont{A.~P.} \bibnamefont{Alivisatos}},
  \bibnamefont{and} \bibinfo{author}{\bibfnamefont{P.~L.}
  \bibnamefont{McEuen}}, \bibinfo{journal}{Nature}
  \textbf{\bibinfo{volume}{407}}, \bibinfo{pages}{57} (\bibinfo{year}{2000}).

\bibitem[{\citenamefont{Grabert and Devoret}(1992)}]{Grabert92}
\bibinfo{author}{\bibfnamefont{H.}~\bibnamefont{Grabert}} \bibnamefont{and}
  \bibinfo{author}{\bibfnamefont{M.~H.} \bibnamefont{Devoret}},
  \emph{\bibinfo{title}{Single Charge Tunneling, NATO ASI Series, Vol.294}}
  (\bibinfo{publisher}{New York, Plenum Press}, \bibinfo{year}{1992}).

\bibitem[{\citenamefont{K\"onig et~al.}(1996)\citenamefont{K\"onig, Schmid,
 Schoeller, and Sch\"on}}]{Koenig96}
\bibinfo{author}{\bibfnamefont{J.} \bibnamefont{K\"onig}},
  \bibinfo{author}{\bibfnamefont{J.} \bibnamefont{Schmid}},
  \bibinfo{author}{\bibfnamefont{H.} \bibnamefont{Schoeller}},
  \bibnamefont{and} \bibinfo{author}{\bibfnamefont{G.}~\bibnamefont{Sch\"on}},
  \bibinfo{journal}{Phys.\ Rev.\ B} \textbf{\bibinfo{volume}{54}},
  \bibinfo{pages}{16820} (\bibinfo{year}{1996}).

\bibitem[{\citenamefont{Hettler et~al.}(2003)\citenamefont{Hettler, Wenzel,
  Wegewijs, and Schoeller}}]{hettler_prl}
  \bibinfo{author}{\bibfnamefont{M.~H.} \bibnamefont{Hettler}},
  \bibinfo{author}{\bibfnamefont{W.} \bibnamefont{Wenzel}},
  \bibinfo{author}{\bibfnamefont{M.~R.} \bibnamefont{Wegewijs}},
  \bibnamefont{and} \bibinfo{author}{\bibfnamefont{H.}~\bibnamefont{Schoeller}},
  \bibinfo{journal}{Phys.\ Rev.\ Lett.} \textbf{\bibinfo{volume}{90}},
  \bibinfo{pages}{076805} (\bibinfo{year}{2003}).

\bibitem[{\citenamefont{Haug and Jauho}(1996)}]{Jauho}
\bibinfo{author}{\bibfnamefont{H.}~\bibnamefont{Haug}} \bibnamefont{and}
  \bibinfo{author}{\bibfnamefont{A.}~\bibnamefont{Jauho}},
  \emph{\bibinfo{title}{Quantum Kinetics in Transport and Optics of
  Semiconductors}} (\bibinfo{publisher}{Springer-Verlag Berlin},
  \bibinfo{year}{1996}).

\bibitem[{\citenamefont{Rammer and Smith}(1986)}]{Rammer86}
\bibinfo{author}{\bibfnamefont{J.}~\bibnamefont{Rammer}} \bibnamefont{and}
  \bibinfo{author}{\bibfnamefont{H.}~\bibnamefont{Smith}},
  \bibinfo{journal}{Rev.\ Mod.\ Phys.} \textbf{\bibinfo{volume}{58}},
  \bibinfo{pages}{323} (\bibinfo{year}{1986}).

\bibitem{note1} The divergence 
of a free sections cutting fermion 
lines associated with states $l$ and $m$ of the same onsite energy arises from the integral 
$\int_{-\infty}^t dt' \exp[-i(\tilde{\epsilon}_m-\tilde{\epsilon}_l)t']$. 
Physically this corresponds to a free propagation for an infinitely long time.

\bibitem[{\citenamefont{Konstantinov and Perel}(1961)}]{Konstantinov61}
\bibinfo{author}{\bibfnamefont{O.~V.} \bibnamefont{Konstantinov}}
  \bibnamefont{and} \bibinfo{author}{\bibfnamefont{V.~I.} \bibnamefont{Perel}},
  \bibinfo{journal}{Sov.\ Phys.\ JETP} \textbf{\bibinfo{volume}{12}},
  \bibinfo{pages}{142} (\bibinfo{year}{1961}).

\bibitem[{\citenamefont{B\"ottger et~al.}(1993)\citenamefont{B\"ottger,
  Bryksin, and Schulz}}]{Boettger93}
\bibinfo{author}{\bibfnamefont{H.}~\bibnamefont{B\"ottger}},
  \bibinfo{author}{\bibfnamefont{V.~V.} \bibnamefont{Bryksin}},
  \bibnamefont{and} \bibinfo{author}{\bibfnamefont{F.}~\bibnamefont{Schulz}},
  \bibinfo{journal}{Phys.\ Rev.\ B} \textbf{\bibinfo{volume}{48}},
  \bibinfo{pages}{161} (\bibinfo{year}{1993}).

\bibitem[{\citenamefont{B\"ottger and Bryksin}(1975)}]{Boettger75}
\bibinfo{author}{\bibfnamefont{H.}~\bibnamefont{B\"ottger}} \bibnamefont{and}
  \bibinfo{author}{\bibfnamefont{V.~V.} \bibnamefont{Bryksin}},
  \bibinfo{journal}{Phys.~Stat.~Sol. B} \textbf{\bibinfo{volume}{71}},
  \bibinfo{pages}{93} (\bibinfo{year}{1975}).

\bibitem[{\citenamefont{B\"ottger and Bryksin}(1976)}]{Boettger76}
\bibinfo{author}{\bibfnamefont{H.}~\bibnamefont{B\"ottger}} \bibnamefont{and}
  \bibinfo{author}{\bibfnamefont{V.~V.} \bibnamefont{Bryksin}},
  \bibinfo{journal}{Phys.~Stat.~Sol. B} \textbf{\bibinfo{volume}{78}},
  \bibinfo{pages}{9} (\bibinfo{year}{1976}).

\bibitem[{\citenamefont{Endres et~al.}(2004)\citenamefont{Endres, Cox, and
  Songh}}]{Endres04}
\bibinfo{author}{\bibfnamefont{R.~G.} \bibnamefont{Endres}},
  \bibinfo{author}{\bibfnamefont{D.~L.} \bibnamefont{Cox}}, \bibnamefont{and}
  \bibinfo{author}{\bibfnamefont{R.~R.~P.} \bibnamefont{Songh}},
  \bibinfo{journal}{Rev.\ Mod.\ Phys.} \textbf{\bibinfo{volume}{76}},
  \bibinfo{pages}{195} (\bibinfo{year}{2004}).

\bibitem[{\citenamefont{Senthilkumar et~al.}(2005)\citenamefont{Senthilkumar,
  Grozema, {Fonseca Guerra}, Bickelhaupt, Lewis, Berlin, Ratner, and
  Siebbeles}}]{Senthilkumar05}
\bibinfo{author}{\bibfnamefont{K.}~\bibnamefont{Senthilkumar}},
  \bibinfo{author}{\bibfnamefont{F.~C.} \bibnamefont{Grozema}},
  \bibinfo{author}{\bibfnamefont{C.}~\bibnamefont{{Fonseca Guerra}}},
  \bibinfo{author}{\bibfnamefont{F.~M.} \bibnamefont{Bickelhaupt}},
  \bibinfo{author}{\bibfnamefont{F.~D.} \bibnamefont{Lewis}},
  \bibinfo{author}{\bibfnamefont{Y.~A.} \bibnamefont{Berlin}},
  \bibinfo{author}{\bibfnamefont{M.~A.} \bibnamefont{Ratner}},
  \bibnamefont{and} \bibinfo{author}{\bibfnamefont{L.~D.~A.}
  \bibnamefont{Siebbeles}}, \bibinfo{journal}{J.\ Am.\ Chem.\ Soc.}
  \textbf{\bibinfo{volume}{127}}, \bibinfo{pages}{14894}
  (\bibinfo{year}{2005}).

\bibitem[{\citenamefont{Alexandre et~al.}(2003)\citenamefont{Alexandre,
  Artacho, Soler, and Chacham}}]{Alexandre03}
\bibinfo{author}{\bibfnamefont{S.~S.} \bibnamefont{Alexandre}},
  \bibinfo{author}{\bibfnamefont{E.}~\bibnamefont{Artacho}},
  \bibinfo{author}{\bibfnamefont{J.~M.} \bibnamefont{Soler}}, \bibnamefont{and}
  \bibinfo{author}{\bibfnamefont{H.}~\bibnamefont{Chacham}},
  \bibinfo{journal}{Phys.\ Rev.\ Lett.} \textbf{\bibinfo{volume}{91}},
  \bibinfo{pages}{108105} (\bibinfo{year}{2003}).

\bibitem[{\citenamefont{Conwell and Rakhmanova}(2000)}]{Conwell00}
\bibinfo{author}{\bibfnamefont{E.~M.} \bibnamefont{Conwell}} \bibnamefont{and}
  \bibinfo{author}{\bibfnamefont{S.}~\bibnamefont{Rakhmanova}},
  \bibinfo{journal}{Proc.\ Natl.\ Acad.\ Sci.\ USA}
  \textbf{\bibinfo{volume}{97}}, \bibinfo{pages}{4556} (\bibinfo{year}{2000}).

\bibitem[{\citenamefont{Henderson et~al.}(1999)\citenamefont{Henderson, Jones,
  Hampikian, Kan, and Schuster}}]{Henderson99}
\bibinfo{author}{\bibfnamefont{P.}~\bibnamefont{Henderson}},
  \bibinfo{author}{\bibfnamefont{D.}~\bibnamefont{Jones}},
  \bibinfo{author}{\bibfnamefont{G.}~\bibnamefont{Hampikian}},
  \bibinfo{author}{\bibfnamefont{Y.}~\bibnamefont{Kan}}, \bibnamefont{and}
  \bibinfo{author}{\bibfnamefont{G.}~\bibnamefont{Schuster}},
  \bibinfo{journal}{Proc.\ Natl.\ Acad.\ Sci.\ USA}
  \textbf{\bibinfo{volume}{96}}, \bibinfo{pages}{8353} (\bibinfo{year}{1999}).

\bibitem[{\citenamefont{Joy et~al.}(2006)\citenamefont{Joy, Guler, Ahmed,
  McLaughlin, and Schuster}}]{Joy06}
\bibinfo{author}{\bibfnamefont{A.}~\bibnamefont{Joy}},
  \bibinfo{author}{\bibfnamefont{G.}~\bibnamefont{Guler}},
  \bibinfo{author}{\bibfnamefont{S.}~\bibnamefont{Ahmed}},
  \bibinfo{author}{\bibfnamefont{L.~W.} \bibnamefont{McLaughlin}},
  \bibnamefont{and} \bibinfo{author}{\bibfnamefont{G.~B.}
  \bibnamefont{Schuster}}, \bibinfo{journal}{Faraday Discuss.}
  \textbf{\bibinfo{volume}{131}}, \bibinfo{pages}{357} (\bibinfo{year}{2006}).

\bibitem[{\citenamefont{triberis}(2005)}]{Triberis05}
\bibinfo{author}{\bibfnamefont{G.~P.} \bibnamefont{Triberis}},
 \bibinfo{author}{\bibfnamefont{C.}~\bibnamefont{Simserides}},
 \bibinfo{author}{\bibfnamefont{V.~C.}~\bibnamefont{Karvolas}},
  \bibinfo{journal}{J.\ Phys.: Condens. Matter} \textbf{\bibinfo{volume}{17}},
  \bibinfo{pages}{2681} (\bibinfo{year}{2005}).

\bibitem[{\citenamefont{triberis}(2009)}]{Triberis09}
\bibinfo{author}{\bibfnamefont{G.~P.} \bibnamefont{Triberis}},
 \bibinfo{author}{\bibfnamefont{M.}~\bibnamefont{Dimakogianni}},
  \bibinfo{journal}{J.\ Phys.: Condens. Matter} \textbf{\bibinfo{volume}{21}},
  \bibinfo{pages}{035114} (\bibinfo{year}{2009}).

\bibitem[{\citenamefont{Starikov}(2005)}]{Starikov05}
\bibinfo{author}{\bibfnamefont{E.~B.} \bibnamefont{Starikov}},
  \bibinfo{journal}{Phil.\ Mag.} \textbf{\bibinfo{volume}{85}},
  \bibinfo{pages}{3435} (\bibinfo{year}{2005}).

\bibitem[{\citenamefont{Olofsson and Larsson}(2001)}]{Olofsson01}
\bibinfo{author}{\bibfnamefont{J.}~\bibnamefont{Olofsson}} \bibnamefont{and}
  \bibinfo{author}{\bibfnamefont{S.}~\bibnamefont{Larsson}},
  \bibinfo{journal}{J.\ Phys.\ Chem.\ B} \textbf{\bibinfo{volume}{105}},
  \bibinfo{pages}{10398} (\bibinfo{year}{2001}).

\end{thebibliography}
\end{document}